\documentclass[runningheads]{llncs}

\usepackage[T1]{fontenc}
\usepackage{graphicx}
\usepackage{amsmath,amssymb}
\usepackage{algorithm2e}
\usepackage{booktabs}
\usepackage{array}
\usepackage{xcolor}
\usepackage{url}
\usepackage{hyperref}
\usepackage{microtype}
\usepackage{multirow}
\usepackage{enumitem}

\SetAlgoLined
\SetKwInOut{Input}{Input}
\SetKwInOut{Output}{Output}
\DontPrintSemicolon

\newcommand{\ABAS}{\textsc{ABAS}}
\newcommand{\BAPDF}{\textsc{bapdf}}
\newcommand{\Jend}{J_{\mathrm{end}}}
\newcommand{\Jopp}{J_{\mathrm{opp}}}
\newcommand{\Ratt}{R_{\mathrm{att}}}
\newcommand{\Renh}{R_{\mathrm{enh}}}
\newcommand{\VA}{V_{\!A}}
\newcommand{\VR}{V_{\!R}}

\begin{document}

\title{Evaluation of
Alternative-Based Information Systems
for Deliberative Polling
using an Agentic Simulator }

\titlerunning{ABAS: Agentic Simulator for Deliberative Polling}

\author{Rwaida Alssadi$^*$, Khulud Alawaji$^*$, Balaji Kasula$^*$, Muntaser Syed$^*$, Badria Alfurhood$^\dagger$, Markus Zanker$^\ddagger$, Marius Silaghi$^*$}
\authorrunning{Alssadi et al.}
\institute{$^*$Florida Institute of Technology\\$^\dagger$Princess Nourah Bint Abdulrahman\\$^\ddagger$Free University of Bozen-Bozano}

\maketitle

\begin{abstract}

Deliberative polling promises to improve collective decision-making by
exposing shareholders to a broad range of arguments before they vote.
Yet ensuring that every voter encounters a representative sample of the
reason space, the \emph{coverage} problem, remains an open challenge,
particularly at scale and in adversarial or strategically motivated
electorates. This paper introduces a way of evaluating solutions using the LLM-based Agentic Bipolar
Argumentation Simulator, 
grounded in a framework
which formalises a poll as a
six-tuple $\langle\Jend,\Jopp,\Ratt,\Renh,\VA,\VR\rangle$ of endorsing
and opposing justifications, attack and enhance relations, and
shareholder- and relation-weights. ABAS simulates $N$ autonomous
shareholder agents, each assigned a latent opinion according to desired distributions in
$[-1,1]$, who sequentially vote, choose or author justifications, and
optionally submit argumentation-graph links. 
The simulator implements
recommendations that rank existing justifications by their \emph{observable} endorsement mass.
It evaluates the mechanism's success by \emph{coverage}, namely the fraction of the corpus reason-tag set represented in the $K$ recommendations presented to each shareholder,
as a solution to the NP-hard Subsuming Justification Problem. 
Reported experiments 
characterise how creativity rate ($p_{\mathrm{own}}$), recommendation size ($K$), argumentation density ($p_{\mathrm{links}}$),
and population size ($N$) affect coverage and corpus diversity.
In an
authenticated electorate where Sybil attacks are impossible and only the
relation graph is gameable, we stress-test the scoring with coordinated strategic voting
attacks: 
a tag-flood attack collapses coverage,
while author-count relation weighting through a reversed-PageRank rule resists the flood markedly better
than uniform weights.
ABAS also supports
repeated interaction in
deliberation, persistent corpus archiving, simulation of strategic opinion presentation, and an interactive
browser for auditability.

\end{abstract}

\section{Introduction}
\label{sec:intro}

Democratic shareholder assemblies' legitimacy increasingly depends not merely on what wins a
vote, but on whether voters were exposed to the full range of relevant
arguments before casting their ballots. Classical democratic theory,
from Mill's marketplace of ideas to Habermas's communicative rationality
\cite{habermas1984theory}, insists that good collective decisions emerge
from genuine deliberation, not from uninformed preferences. Deliberative
polling, systematically developed by Fishkin~\cite{fishkin1991democracy},
operationalises this ideal by measuring opinion change after structured
exposure to diverse arguments. Empirical deployments consistently show
that access to a broader argumentative corpus improves decision quality
and reduces polarisation \cite{fishkin2009deliberative}.
The {\bf Alternative-Based Information System (ABIS)} debate structure model for simultaneously informing and letting shareholder decide in authenticated shareholder organisations was introduced in~\cite{silaghi2017they}.

The digital era has transformed the scale at which deliberations
can operate. Online platforms such as Polis \cite{small2023polis},
LiquidFeedback, and Kialo aggregate hundreds of thousands of
contributions to inform organisers; AI-mediated tools can now assist with argument generation,
clustering, and recommendation. Yet scale introduces a tension. As the
corpus of justifications grows, the probability that any single voter
encounters a representative sample of the reason space shrinks. This is
the \emph{coverage problem}: how should a recommendation engine select
at most $K$ justifications for each voter such that, taken together,
the recommendations presented across the entire electorate faithfully
represent the full argument space?

The coverage problem is not merely a usability concern. Systematic
coverage gaps introduce \emph{epistemic injustice}: voters who vote
early in a sequential process see sparse recommendations drawn from a
small pool, while late voters benefit from a mature corpus. Polarisation
and echo-chamber dynamics \cite{pariser2011filter,sunstein2002republic}
can further distort coverage if recommendation algorithms optimise for
engagement rather than epistemic breadth. 

Former evaluations of mechanism impact on achieved coverage needed complex human subjects interviews and panels that produced hard to reproduce and verify results.
Formal, reproducible evaluation
of coverage under varying behavioural assumptions requires a simulation
infrastructure that so far did not exist and was only enabled by the advent of LLM technology~\cite{argyle2023outofone,park2023generative,wu2023autogen}.
This paper addresses that gap. We present a way to evaluate solutions via LLM-based \ABAS{} (Agentic Bipolar
Argumentation Simulator), a
simulator grounded in
a framework for reasoning about the comprehensiveness of justification
sets in bipolar argumentation contexts, including the usable definition
of the NP-hard Subsuming Justification Problem (SJP)~\cite{silaghi2017bapdf}.
\ABAS{} gives this framework computational life: it
instantiates $N$ autonomous shareholder agents, simulates their voting
and argumentation behaviour across configurable probability parameters,
and measures the resulting coverage and diversity of the recommendation
snapshots each agent receives.

As a contribution we introduce an \textbf{Agentic Bipolar Argumentation Simulator (\ABAS{}).} This is a
    seed-controlled simulator
    implementation of the Bipolar Abstract Poll Debate Framework (\BAPDF{}) deliberative
    poll model
    with trace persistence, 
    endorsement-ranked recommendation evaluated
    against a greedy-coverage oracle,
    and multi interaction support.
This simulator enables us to perform  \textbf{sensitivity analysis} on mechanism nuances. 
A systematic 400-experiment
    study (two topics) examines the effects of creativity rate ($p_{\mathrm{own}}$),
    recommendation size ($K$), argumentation density ($p_{\mathrm{links}}$),
    and population size ($N$) on coverage and corpus diversity.
  To study the impact of data availability on early participants, the simulator has a \textbf{multi-round deliberation design.} This works as a deactivate-and-rerun
    protocol that preserves the justification corpus across rounds for
    longitudinal auditability.
For user interface evaluations and/or debugging situations, an  \textbf{interactive browser} is also provided. This is a graphical interface exposing
    per-shareholder recommendation snapshots, transitive argumentation
    coverage visualisations, and relation graphs.

Section~\ref{sec:related} reviews
related work on deliberative polling, abstract argumentation, and
argument recommendation. Section~\ref{sec:bapdf} introduces the \BAPDF{}
model and the coverage metric. Section~\ref{sec:abis} describes the \ABAS{}
simulator architecture. Section~\ref{sec:setup} presents the experimental
setup. Section~\ref{sec:results} reports results. Section~\ref{sec:discussion}
discusses implications, and Section~\ref{sec:conclusion} concludes.

\section{Related Work}
\label{sec:related}

The theoretical foundations of deliberative democracy trace to
Habermas's \cite{habermas1984theory} ideal of communicative rationality:
legitimate collective decisions emerge from discourse in which all
affected parties may contribute arguments on equal terms. Fishkin
\cite{fishkin1991democracy,fishkin2009deliberative} operationalised this
ideal as \emph{deliberative polling}, in which a stratified random sample
of citizens deliberates over briefing materials before and after
structured discussion. Repeated experiments spanning dozens of countries
demonstrate that deliberation shifts opinions toward greater consistency
and reduces the influence of party cues, but these experiments typically
involve small cohorts (200–500 participants) and facilitated face-to-face
discussion. Scaling deliberative polling to national or global electorates
requires computational mediation.

\subsection{Abstract Argumentation Frameworks}

Dung's \cite{dung1995acceptability} seminal framework models
argumentation as a directed graph in which arguments attack one another,
and defines several \emph{extension semantics} (stable, preferred,
grounded) that identify collectively acceptable subsets of arguments.
This work ignited a large literature on computational argumentation
\cite{bench2007argumentation,baroni2018introduction}, with applications
in legal reasoning, multi-agent negotiation, and decision support.

Cayrol and Lagasquie-Schiex \cite{cayrol2005bipolarity,cayrol2009bipolarsemantic}
introduced \emph{bipolar} argumentation frameworks by adding a
support/enhance relation alongside
Dung's attack relation. Support captures the observation that arguments
frequently reinforce, rather than merely oppose, one another. Bipolar
frameworks subsume Dung's model and allow richer semantic analyses of
argument interaction. The \BAPDF{} model 
applies bipolarity to the polling domain,
distinguishing endorsing justifications ($\Jend$, favouring the
proposition) from opposing justifications ($\Jopp$, opposing it), and
equipping both sets with attack and enhance relations that cross or
traverse sides, respectively.

\subsection{Argument Mining and Automated Argumentation}

Argument mining \cite{lippi2016argumentation,stab2017parsing} extracts
argumentative structures from free text, enabling downstream applications
in automated counterargument retrieval \cite{wachsmuth2018retrieval},
evaluative argument generation \cite{carenini2006argumentative}, and
persuasive dialogue systems \cite{chalaguine2020argumentation}. These
systems construct argument graphs from corpora and can in principle be
integrated with recommenders to present human deliberators with relevant
and diverse arguments. \ABAS{} currently uses symbolic reason tags as
proxies for argument identities; the integration of argument-mining
pipelines to populate the tag vocabulary from free text is a natural
extension.

\subsection{AI-Assisted Democracy Platforms}

Commercial and academic platforms have explored AI mediation in civic
discourse. Polis \cite{small2023polis} uses dimensionality reduction to
cluster voter opinions and surface areas of consensus; LiquidFeedback
supports liquid-democracy delegation chains; Kialo provides tree-structured
debate with manual pro/con tagging. These systems prioritise usability
and engagement but do not formally evaluate coverage of the argument
space. \ABAS{} complements these platforms by providing a principled
theoretical basis, \BAPDF{}, and systematic empirical evaluation of
coverage properties.

\subsection{Diversity and Coverage in Recommender Systems}

The tradeoff between relevance and diversity in recommender systems has
been extensively studied. Carbonell and Goldstein's \cite{carbonell1998mmr}
Maximal Marginal Relevance formulation explicitly penalises redundancy.
Burke \cite{burke2017multisided} frames fairness in recommendations as a
multisided optimisation. Chakraborty et al.\ \cite{chakraborty2017fairness}
examine diversity in social media recommendation. The \BAPDF{} notion of
\emph{coverage} provides a domain-specific diversity criterion anchored
in formal argumentation theory: a set of justifications is diverse if and
only if it collectively represents the full vocabulary of atomic reasons
that appear in the corpus. This is stronger than generic diversity
because it is tied to a formal definition of argumentative comprehensiveness
(Definition~\ref{def:comprehensiveness}).

\subsection{Agentic Simulation of Social Behaviour}

Recent work has demonstrated that large language models can act as
simulacra of human behaviour at scale. Park et al.\ \cite{park2023generative}
simulate communities of agents with individual memories and social
interactions. Argyle et al.\ \cite{argyle2023outofone} show that GPT-4
can replicate survey distributions across demographic strata. Wu et al.~\cite{wu2023autogen} provide infrastructure for multi-agent LLM
conversations. \ABAS{} contributes to this tradition by providing a
\emph{formal, domain-specific} agent model grounded in argumentation
theory (with configurable behavioural parameters rather than free-form
LLM generation) enabling controlled, extensive sensitivity analyses
that would be infeasible with stochastic LLM-based agents.

\section{The \BAPDF{} Model}
\label{sec:bapdf}

A \emph{deliberative poll} over a binary proposition is modelled in
\BAPDF{} as a six-tuple:
\begin{equation}
  \Pi = \langle\Jend,\, \Jopp,\, \Ratt,\, \Renh,\, \VA,\, \VR\rangle
  \label{eq:bapdf}
\end{equation}
where each component is defined as follows.

\begin{definition}[BAPDF Components]
\label{def:bapdf}
Let $\mathcal{J} = \Jend \cup \Jopp$ be a finite set of
\emph{justifications}, where $\Jend$ (the \emph{endorsing} set) contains
natural-language arguments in favour of the proposition, and $\Jopp$
(the \emph{opposing} set) contains arguments against it.
\begin{itemize}[leftmargin=*,itemsep=1pt]
  \item $\Ratt \subseteq \Jend \times \Jopp \cup \Jopp \times \Jend$
    is the set of \emph{attack} (reply) relations between justifications
    on opposite sides of the debate.
  \item $\Renh \subseteq \Jend \times \Jend \cup \Jopp \times \Jopp$
    is the set of \emph{enhance} (improve) relations between justifications
    on the same side of the debate.
  \item $\VA\colon \mathcal{A} \to \mathbb{R}_{>0}$ is the weight function
    over shareholders $\mathcal{A}$, representing their relative voting
    power.
  \item $\VR\colon \Ratt \cup \Renh \to \mathbb{R}_{>0}$ is the weight
    function over relations, representing the strength of argumentative
    interactions.
\end{itemize}
\end{definition}

{In the \ABAS{} experiments reported here we keep uniform
shareholder weights ($\VA \equiv 1$) but study \emph{both} relation-weighting
regimes. In the so-called \emph{uniform} mode, $\VR \equiv 1$; in \emph{weighted} mode each
relation's weight is proportional to the fraction of side-voters who authored it,
$\VR(r) = |\mathrm{authors}(r)| / |\mathrm{voters}_{\mathrm{side}}|$. The
relation weights feed the endorsement-based scoring of
Section~\ref{sec:scoring}, and the choice between the two regimes is the lever
that distinguishes manipulation-resistant from manipulable configurations
under the adversarial experiments of Section~\ref{sec:res_attackers}.} The bipolar structure allows the argumentation graph to
represent both disagreement (attack) and alignment (enhance) between
arguments, mirroring the structure of real debates in which some
arguments reinforce others on the same side while challenging adversarial
positions.

\subsection{Justification Space and Reason Tags}

Each justification $j \in \mathcal{J}$ is associated with a non-empty
set of \emph{reason tags} $\tau(j) \subseteq \mathcal{T}$, where
$\mathcal{T}$ is a fixed vocabulary of atomic argument identifiers. In
the \ABAS{} implementation, $\mathcal{T} = \mathcal{T}_{Y} \cup \mathcal{T}_{N}$
with $\mathcal{T}_{Y} = \{\mathtt{Y01},\ldots,\mathtt{Y30}\}$ (endorsing
reasons) and $\mathcal{T}_{N} = \{\mathtt{N01},\ldots,\mathtt{N30}\}$
(opposing reasons). Each tag corresponds to a distinct semantic position
on the debate topic. For example, for the test proposition ``Should one address the AI
revolution by introducing basic income?'', $\mathtt{Y01}$ represents
the argument from projected AI-driven mass unemployment, $\mathtt{Y04}$
represents the redistribution of AI productivity gains, $\mathtt{N01}$
represents the historical job-creation counterargument, and so on.
Justification texts are 100--200 word natural-language passages that
instantiate one or more of these atomic reasons.

The reason-tag vocabulary provides a discrete semantic skeleton over the
justification corpus. Unlike free-text embeddings, tags are transparent,
enumerable, and directly amenable to set-theoretic coverage analysis.

\subsection{Coverage Metric}

\begin{definition}[Corpus Tag Set]
\label{def:corpustags}
For side $s \in \{\mathrm{end},\mathrm{opp}\}$, the \emph{corpus tag
set} at time $t$ is:
\[
  \mathcal{T}^s_t = \bigcup_{j \in \mathcal{J}^s_t} \tau(j)
\]
where $\mathcal{J}^s_t$ is the set of justifications on side $s$ that
are active at time $t$.
\end{definition}

\begin{definition}[Recommendation Coverage]
\label{def:coverage}
Let $R_i^s$ be the set of at most $K$ justifications recommended to
shareholder $i$ on side $s$, computed at the time of their vote (so
$\mathcal{T}^s_{t_i}$ is the corpus tag set at that moment). The
\emph{coverage} of the recommendation for shareholder $i$ on side $s$
is:
\[
  c_i^s = \frac{\left|\bigcup_{j \in R_i^s} \tau(j)\right|}{|\mathcal{T}^s_{t_i}|}
\]
The \emph{mean combined coverage} over all shareholders and sides is:
\[
  \bar{c} = \frac{1}{2N} \sum_{i=1}^{N}
            \left( c_i^{\mathrm{end}} + c_i^{\mathrm{opp}} \right)
\]
\end{definition}

This metric captures the fraction of the 
contemporaneous
argument space
that each shareholder's recommendation set represents. A value of $1.0$
indicates that the $K$ recommendations together cover every distinct
reason in the corpus at the time of that shareholder's vote.

A primary limitation of this study lies in the potential thematic overlap between the rationale for support and opposition. The valence of a given statement, i.e., whether it is interpreted as supportive or opposing, is inherently contingent upon the latent values and cognitive constraints of the voter.

\subsection{The Subsuming Justification Problem}

\begin{definition}[Subsuming Justification Set]
\label{def:subsuming}
Given a set of justifications $\mathcal{J}^s$ on side $s$ with tag function
$\tau$, a \emph{$K$-subsuming set} is any $S \subseteq \mathcal{J}^s$
with $|S| \leq K$ such that $\bigcup_{j \in S} \tau(j) = \mathcal{T}^s$.
\end{definition}

\begin{definition}[Comprehensiveness]
\label{def:comprehensiveness}
A recommendation set $R_i^s$ is \emph{comprehensive} with respect to
$\mathcal{J}^s$ if it is a $K$-subsuming set, i.e., $c_i^s = 1$.
\end{definition}

The \emph{Subsuming Justification Problem} (SJP) deciding whether a
$K$-subsuming set exists for given $\mathcal{J}^s$, $\tau$, and $K$, is
NP-hard, as shown by reduction from Minimum Set Cover.

The NP-hardness of SJP motivates our use of greedy approximation.

\subsection{Greedy $K$-Coverage Approximation}

\begin{definition}[Greedy Coverage Algorithm]
\label{def:greedy}
Given $\mathcal{J}^s$ and budget $K$, the greedy algorithm initialises
$S_0 = \emptyset$ and iteratively selects:
\[
  j^* = \arg\max_{j \in \mathcal{J}^s \setminus S_{k-1}}
        \left|\tau(j) \setminus \textstyle\bigcup_{j' \in S_{k-1}} \tau(j')\right|
\]
adding $j^*$ to $S_k$ until $|S_k| = K$ or no further coverage gain is
possible.
\end{definition}

By the submodularity of the set-union function and Nemhauser et al.'s
\cite{nemhauser1978analysis} classical result, this greedy algorithm
achieves a $(1 - 1/e) \approx 0.632$ approximation ratio relative to
the optimal $K$-coverage in polynomial time. In practice, the
approximation is near-optimal when the tag sets of individual
justifications are small relative to $|\mathcal{T}^s|$, a condition
that holds in our experimental corpus.

\subsection{Attack and Enhance Semantics}

Attack relations $\Ratt$ capture argumentative rebuttal: if $(j_1, j_2)
\in \Ratt$ with $j_1 \in \Jopp$ and $j_2 \in \Jend$, then $j_1$
challenges the reasoning of $j_2$. In \ABAS{}, attack targets are
selected by maximising a \emph{tag-attack score}:
\[
  \sigma_{\mathrm{att}}(j_1, j_2) =
    \frac{|\{n : \mathtt{Y}n \in \tau(j_2) \wedge \mathtt{N}n \in \tau(j_1)\}|}
         {|\{n : \mathtt{Y}n \in \tau(j_2)\} \cup \{n : \mathtt{N}n \in \tau(j_1)\}|}
\]
This score is high when the NO justification addresses the same thematic
index numbers as the YES justification, reflecting direct topical
opposition.

Enhance relations $\Renh$ capture argumentative reinforcement: if $(j_1,
j_2) \in \Renh$ with $j_1, j_2 \in \Jend$, then $j_1$ supports or
elaborates $j_2$. Enhance targets are selected by Jaccard similarity:
\[
  \sigma_{\mathrm{enh}}(j_1, j_2) =
    \frac{|\tau(j_1) \cap \tau(j_2)|}{|\tau(j_1) \cup \tau(j_2)|}
\]

\subsection{Transitive Argumentation Coverage}

We will study the evaluation of a metric called \emph{transitive argumentation coverage} that we use
to measure the reach of individual justifications through the
argumentation graph:
\begin{definition}[Transitive Argumentation Coverage]
\label{def:trans}
For justification $j \in \mathcal{J}$, let $\mathcal{A}(j)$ be the set
of shareholders reachable from $j$ by following any sequence of enhance
and attack edges in the argumentation graph. The transitive argumentation
coverage of $j$ is:
\[
  \rho(j) = \frac{|\mathcal{A}(j)|}{|\mathcal{A}|}
\]
where $|\mathcal{A}|$ is the total shareholder count.
\end{definition}

High $\rho(j)$ indicates that justification $j$ is influential across
a large fraction of the electorate, either directly (many shareholders
cited it) or indirectly (it is connected to widely cited justifications
via the argumentation graph). The \ABAS{} browser renders $\rho(j)$ as
a bar chart alongside each justification.
Because the live recommender ranks justifications by an
endorsement score that propagates along enhance and attack relations
(Section~\ref{sec:scoring}), the relation graph is an attack surface for
strategic voting on visibility; we quantify this surface, and the protective
role of non-uniform relation weights, in Section~\ref{sec:res_attackers}.
\section{The \ABAS{} Simulator}
\label{sec:abis}

The simulation stores a complete recommendation snapshot for every
shareholder before their vote is cast, enabling after-the-fact audit
of what information each voter saw at the exact moment of their decision.

\subsection{Shareholder Model}

Each shareholder $a_i \in \mathcal{A}$ is characterised by:
\begin{itemize}[leftmargin=*,itemsep=1pt]
  \item A latent \emph{opinion} $o_i \sim \mathrm{Uniform}(-1, 1)$ (other distributions can also be studied).
  \item A 200--300 word \emph{position text} $p_i$ generated
    deterministically from $o_i$ by selecting opinion-quartile-appropriate
    reason sentences from the tag vocabulary.
  \item A \emph{vote direction}: YES if $o_i > 0$, NO if $o_i < 0$.
\end{itemize}
The position text $p_i$ serves as the query for Term Frequency-Inverse Document Frequency (TF-IDF) similarity
computation when the shareholder selects an existing justification. The
uniform opinion distribution in this study ensures an approximately balanced electorate.

\subsection{Simulation Loop}

Algorithm~\ref{alg:sim} presents the core simulation loop. Shareholders
are processed in ascending ID order. The simulator maintains a live
in-memory index of active justifications, refreshed every
$\Delta = 200$ shareholders or immediately after a new own justification
is added (the \emph{dirty-cache} mechanism). This design ensures that
each shareholder's recommendation is based on the most current corpus
state while avoiding $O(N^2)$ full-rebuild overhead.

\begin{algorithm}[t]
\caption{ABAS Simulation Loop for a Round}\label{alg:sim}
\Input{Shareholders $\mathcal{A}$; probabilities $p_{\mathrm{own}}, p_{\mathrm{existing}},
  p_{\mathrm{none}}$ (sum to 1); link probability $p_{\mathrm{links}}$;
  recommendation budget $K$; round $r$}
\Output{Populated \texttt{votes}, \texttt{recommendations}, \texttt{relations} tables}
Load active justification pools $\mathcal{J}^{\mathrm{end}}_r,
  \mathcal{J}^{\mathrm{opp}}_r$\;
$\mathit{dirty} \leftarrow \mathbf{false}$\;
\For{each shareholder $a_i$ in order}{
  \If{$i \bmod \Delta = 0$ \textbf{or} $\mathit{dirty}$}{
    Refresh $\mathcal{J}^{\mathrm{end}}_r, \mathcal{J}^{\mathrm{opp}}_r$ from DB\;
    $\mathit{dirty} \leftarrow \mathbf{false}$\;
  }
  {$R^{\mathrm{end}}_i,\, R^{\mathrm{opp}}_i \leftarrow \textsc{EndorsementRank}(\mathcal{J}^{\mathrm{end}}_r, \mathcal{J}^{\mathrm{opp}}_r, E, \VR, K)$}\;
  Store snapshot $(a_i, R^{\mathrm{end}}_i, R^{\mathrm{opp}}_i, r)$ in \texttt{recommendations}\;
  $v \leftarrow \mathbf{YES}$ if $o_i > 0$ else $\mathbf{NO}$\;
  $\mathcal{R}_i \leftarrow R^{\mathrm{end}}_i$ if $v = \mathbf{YES}$ else $R^{\mathrm{opp}}_i$\;
  $u \leftarrow \mathrm{Uniform}(0,1)$\;
  \uIf{$u < p_{\mathrm{existing}}$ \textbf{and} $\mathcal{R}_i \neq \emptyset$}{
    $j_i \leftarrow \textsc{SelectBySimilarity}(p_i, \mathcal{R}_i, v)$\;
  }
  \uElseIf{$u < p_{\mathrm{existing}} + p_{\mathrm{none}}$}{
    $j_i \leftarrow \mathbf{null}$\;
  }
  \Else{
    $(j_i, \tau_i) \leftarrow \textsc{GenerateOwnJustification}(p_i, v)$\;
    Insert $j_i$ into $\mathcal{J}^v_r$ and TF-IDF index\;
    $\mathit{dirty} \leftarrow \mathbf{true}$\;
  }
  Insert vote $(a_i, v, j_i, r)$ into \texttt{votes}\;
  \If{$j_i \neq \mathbf{null}$ \textbf{and} $\mathrm{Uniform}(0,1) < p_{\mathrm{links}}$}{
    $j^{\mathrm{att}} \leftarrow \textsc{FindMostAttacked}(\tau(j_i), \mathcal{J}^{\bar{v}}_r)$\;
    $j^{\mathrm{enh}} \leftarrow \textsc{FindClosestSupporting}(\tau(j_i), \mathcal{J}^v_r)$\;
    Insert relations $(j_i \to j^{\mathrm{att}}, \mathrm{attack})$ and
      $(j_i \to j^{\mathrm{enh}}, \mathrm{enhance})$ if found\;
  }
}
\end{algorithm}

\subsection{Recommendation of Justifications}
\label{sec:scoring}

{The recommendation mechanism combines two strategies: an
\emph{endorsement-based ranking} that generates the shared top-$K$ list shown
to every voter, and Term Frequency-Inverse Document Frequency (\emph{TF-IDF similarity}) for the individual justification a
voter then reuses. We assume that an acceptable, deployable recommender cannot objectively read the oracle reason-tags
$\tau(j)$ (AI attempts to guess it could be accused as systematic manipulation vectors); it can only observe objectively which justifications shareholders have endorsed.
Greedy $K$-coverage (Algorithm~\ref{alg:greedy}) is therefore retained as an
\emph{oracle upper bound} for evaluation, never on the live path, and the
coverage we report measures how closely the observable endorsement ranking
approximates that oracle.}

\paragraph{
Endorsement-based scoring (reversed PageRank).}
{Let $E(j)$ be the number of shareholders who endorsed
justification $j$ (i.e.\ cast their vote citing $j$). The engine ranks the
side-$s$ pool $\mathcal{J}^s_r$ by a score in which a source inherits a share
of the endorsement mass of the justifications it points to:}
\begin{equation}
  {\mathrm{score}(j) = E(j)
    + \!\!\sum_{(j,k)\in\Renh}\!\! \VR(j,k)\,E(k)
    + \!\!\sum_{(j,k')\in\Ratt}\!\! \VR(j,k')\,E^{\mathrm{opp}}(k')}
  \label{eq:score}
\end{equation}
{where the enhance term launders same-side endorsements and the
attack term makes rebuttals of salient opposing arguments more visible
($E^{\mathrm{opp}}$ being opposite-side endorsements). This is a one-hop
reversed-PageRank: endorsement mass flows backward along relations to their
authors, weighted by $\VR$ (ideally this would be multi-hop but the weights tend to be small and remote hops anyhow vanish). In uniform mode ($\VR\equiv1$) a single relation
transfers the target's full endorsement count; in weighted mode
($\VR=|\mathrm{authors}|/|\mathrm{voters}_{\mathrm{side}}|$) the transfer is
throttled by the fraction of the side that backs the relation. The top-$K$
list is cached and refreshed every $\Delta$ shareholders or when a new own
justification is added.}

\paragraph{Greedy $K$-Coverage 
(evaluation oracle).}
Given the current pool $\mathcal{J}^s_r$, 
the oracle applies
Algorithm~\ref{alg:greedy} to select the $K$ justifications that
maximise $\bigcup_{j \in S} \tau(j)$. 
It defines the best
attainable coverage for budget $K$; the gap between it and the endorsement
ranking is the price of using only observable signals. Caching is
essential for performance: a naive per-shareholder greedy computation
over a pool of $|\mathcal{J}| \approx 120$ justifications with $N = 1{,}000$
shareholders would require $O(N \cdot |\mathcal{J}| \cdot K)$ tag-set
operations; caching reduces this to $O(|\mathcal{J}| \cdot K)$ amortised.

\begin{algorithm}[t]
\caption{Greedy K-Coverage Recommendation}\label{alg:greedy}
\Input{Justification pool $\mathcal{J}^s$; budget $K$}
\Output{Ordered recommendation list $S$ of size $\leq K$}
$S \leftarrow \emptyset$; $C \leftarrow \emptyset$
  \tcp*[r]{selected set; covered tags}\;
\While{$|S| < K$ \textbf{and} $\mathcal{J}^s \setminus S \neq \emptyset$}{
  $j^* \leftarrow \arg\max_{j \in \mathcal{J}^s \setminus S} |\tau(j) \setminus C|$\;
  $S \leftarrow S \cup \{j^*\}$;
  $C \leftarrow C \cup \tau(j^*)$\;
}
\Return $S$\;
\end{algorithm}

\paragraph{TF-IDF Similarity Selection.}
When a shareholder chooses to reuse an existing justification (with
probability $p_{\mathrm{existing}}$), the engine selects from the
recommendation list $\mathcal{R}_i$ using TF-IDF cosine similarity
between the shareholder's position text $p_i$ and each justification's
text. A TF-IDF vectoriser with bigrams and sublinear term-frequency
scaling is maintained incrementally; the selection is drawn from a
softmax distribution over similarity scores to avoid determinism that could reduce experimental alternatives.

\subsection{Database Schema}\sloppy

A database persists the complete simulation state with six tables.
The \texttt{justifications} table stores the text, side, reason-tag JSON
array, optional author (shareholder) ID, round number, and active flag.
The \texttt{votes} table records each shareholder's vote direction and
chosen justification. The \texttt{relations} table records attack and
enhance relations with author, weight, round, and active flag. The
\texttt{recommendations} table stores the ranked $K$-justification
snapshot presented to each shareholder, enabling exact reconstruction
of what each voter saw. The \texttt{simulation\_params} table records
all simulation parameters for reproducibility.

The multi-round design deactivates votes and relations from previous
rounds (preserving them for audit) while retaining the justification
corpus, which accumulates across rounds. Justifications submitted in
earlier rounds remain visible and recommendable in later rounds, modelling
the persistent collective memory of a deliberative body.

The \ABAS{} browser provides a paginated interface to the
simulation database. For each shareholder it displays the opinion, position text
and vote; the recommendation snapshot stored at vote time; the transitive
argumentation coverage $\rho(j)$ of each recommended justification; and relation
chips for the attack and enhance links touching that shareholder's justification.
By supporting both at-vote-time and current-state recommendation views, it lets
auditors compare the information a shareholder had against the current corpus.

\section{Experimental Setup}
\label{sec:setup}

{All experiments are run on \emph{two} real-world propositions,
each with its own hand-authored reason-tag vocabulary and text generator:
\textbf{BRA} --- \emph{``Should one address the AI revolution by introducing
basic resource assurance (minimal healthy housing, food, and emergency
healthcare)?''} --- and \textbf{UBI} --- \emph{``Should one address the AI
revolution by introducing basic income?''} Both engage a wide range of
economic, social, and political arguments and were chosen because their thirty
YES reason tags ($\mathtt{Y01}$--$\mathtt{Y30}$) and thirty NO reason tags
($\mathtt{N01}$--$\mathtt{N30}$) each span distinct argument dimensions,
providing a rich but enumerable argument space. The two topics yield nearly
identical sensitivity behaviour, so we present figures for BRA and report UBI
alongside it in the consolidated tables.} Importantly, each shareholder's vote is entirely
determined by their latent opinion and not by the content of recommended
justifications, ensuring that we measure the informational completeness
of the recommendation process without introducing feedback loops.

\subsection{Parameter Space}

We conduct a one-factor-at-a-time sensitivity analysis across four
parameters, holding all others at their baseline values:

\begin{itemize}[leftmargin=*,itemsep=2pt]
  \item \textbf{Creativity rate} $p_{\mathrm{own}} \in \{0.02, 0.05,
    0.10, 0.20, 0.30\}$, with $p_{\mathrm{existing}} = 0.50$ and
    $p_{\mathrm{none}} = 1 - p_{\mathrm{own}} - p_{\mathrm{existing}}$
    adjusted to maintain the probability simplex.
  \item \textbf{Recommendation size} $K \in \{5, 10, 15, 20, 30\}$.
  \item \textbf{Link probability} $p_{\mathrm{links}} \in \{0.20, 0.40,
    0.60, 0.80, 1.00\}$.
  \item \textbf{Population size} $N \in \{200, 500, 1{,}000, 2{,}000,
    5{,}000\}$.
\end{itemize}

The baseline configuration is $p_{\mathrm{own}} = 0.10$,
$p_{\mathrm{existing}} = 0.50$, $p_{\mathrm{none}} = 0.40$,
$p_{\mathrm{links}} = 0.60$, $K = 20$, $N = 1{,}000$. Each configuration
is run with 10 random seeds, yielding $5 \times 4
\times 10 = 200$ simulation runs 
per topic ($400$ in total).
All sensitivity runs use the weighted relation mode.

\subsection{
Adversarial Model}
\label{sec:setup_attackers}

{Because \ABAS{} assumes an \emph{authenticated} electorate,
each identity casts exactly one honest vote and one honest endorsement, so
Sybil (clone-identity) attacks are impossible by construction. The only
gameable surface is the \emph{relation graph}, which feeds the endorsement
score of Eq.~\ref{eq:score}. We model a coordinated coalition of
$s_a \in \{0,50,100,150,200,250\}$ strategic shareholders (0\%--25\% of $N$)
who still vote sincerely but choose their relations adversarially, and study
two strategies:}
\begin{itemize}[leftmargin=*,itemsep=1pt]
  \item \textbf{
  Hub-riding}: 
  each attacker
    submits an enhance relation from its own justification to the most-endorsed
    same-side justification (the \emph{hub}), laundering $E_{\mathrm{hub}}$ into
    its own score (Eq.~\ref{eq:score}) to climb the ranking.
  \item \textbf{
  Tag-flood}: 
  a dual-hub variant
    in which every attacker authors a justification carrying the \emph{same}
    narrow two-tag ``poison'' set, enhances the same-side hub, and attacks the
    most-endorsed opposite-side hub. Tag-homogeneous clones thus ride the hubs'
    endorsement mass into the top-$K$ and evict tag-diverse slots.
\end{itemize}
{Attacker arrival order is a confound, so attackers are
distributed uniformly across the processing order (\emph{spread} timing), which
exposes the relation lever without granting a first-mover endorsement cascade.
Each attacker cell is run with 200 seeds for both relation-weighting modes and
both topics.}

\subsection{Metrics}

We report the following metrics aggregated across seeds:
\begin{itemize}[leftmargin=*,itemsep=1pt]
  \item \textbf{Mean combined coverage} $\bar{c}$ and its standard
    deviation $\sigma_c$ across shareholders.
  \item \textbf{Median combined coverage}: robust to the heavy left tail
    caused by early shareholders who vote before the corpus matures.
  \item \textbf{Own justification count}: total number of shareholder-authored
    justifications, split by YES/NO side, a proxy for corpus diversity.
  \item \textbf{Unique reason tags}: number of distinct tags appearing
    in the corpus, also split by side.
  \item \textbf{Relation counts}: total attack and enhance relations,
    a measure of argumentation graph density.
\end{itemize}

\subsection{Hardware and Software}

Experiments were executed on a macOS 14 system 
with 16 GB RAM. 
Each 1,000-shareholder simulation completes in approximately 18--45 seconds depending on the parameter configuration; 
the full 400-experiment sensitivity suite (two topics), together with the adversarial sweeps, required several hours of wall-clock time.

\section{Results}
\label{sec:results}

Table~\ref{tab:baseline} summarises the baseline configuration
($p_{\mathrm{own}} = 0.10$, $K = 20$, $p_{\mathrm{links}} = 0.60$,
$N = 1{,}000$) for the BRA topic across 10 seeds. 
The mean
combined coverage of the endorsement-ranked recommendations is $0.808$
with a standard deviation of $0.035$ across seeds, indicating moderate
run-to-run variability. The median combined coverage of $0.880$
corroborates the absence of severe outliers at the aggregate level; UBI
behaves almost identically ($0.789 \pm 0.031$).

The coverage standard deviation \emph{across shareholders within a
single run}
is
substantially larger ({approximately $0.19$}) reflecting the early-arrival
penalty: shareholders who vote before the corpus matures receive sparse
recommendations, while late-arriving shareholders reliably receive
comprehensive lists. This within-run variance is a structural property
of sequential deliberation and can be mitigated by increasing $p_{\mathrm{own}}$
or $N$ so that the corpus matures earlier.

At baseline, each run produces approximately $97$ own
justifications and $687$ argumentation edges, and all 30 YES and (almost) all 30
NO reason tags are represented by the end of every run (Table~\ref{tab:baseline}).

\begin{table}[t]
\centering
\caption{
Baseline configuration statistics for the BRA topic
  ($p_{\mathrm{own}}=0.10$,
  $K=20$, $p_{\mathrm{links}}=0.60$, $N=1{,}000$; mean $\pm$ std over 10 seeds).
  Coverage is that of the endorsement-ranked top-$K$ list; within-run std refers
  to the standard deviation of per-shareholder coverage scores within
  a single run (averaged across 10 seeds).}
\label{tab:baseline}
\setlength{\tabcolsep}{6pt}
\begin{tabular}{lrr}
\toprule
\textbf{Metric} & \textbf{Mean (across seeds)} & \textbf{Std (across seeds)} \\
\midrule
Combined coverage (mean)     & 0.808 & 0.035 \\
Combined coverage (median)   & 0.880 & 0.045 \\
Within-run coverage std      & 0.193 & 0.024 \\
YES tags in corpus           & 30.0 & 0.0 \\
NO tags in corpus            & 29.8  & 0.4   \\
Own YES justifications       & 49.4 & 8.5 \\
Own NO justifications        & 48.4 & 5.9 \\
Attack relations             & 345 & 18 \\
Enhance relations            & 343 & 15 \\
Total relations              & 687 & 32 \\
YES votes                    & 495.3 & 18.2 \\
NO votes                     & 504.7 & 18.2 \\
\bottomrule
\end{tabular}
\end{table}

\subsection{Effect of Creativity Rate ($p_{\mathrm{own}}$)}
\label{sec:res_pown}

\begin{figure}[t]
  \centering
  \includegraphics[width=\linewidth]{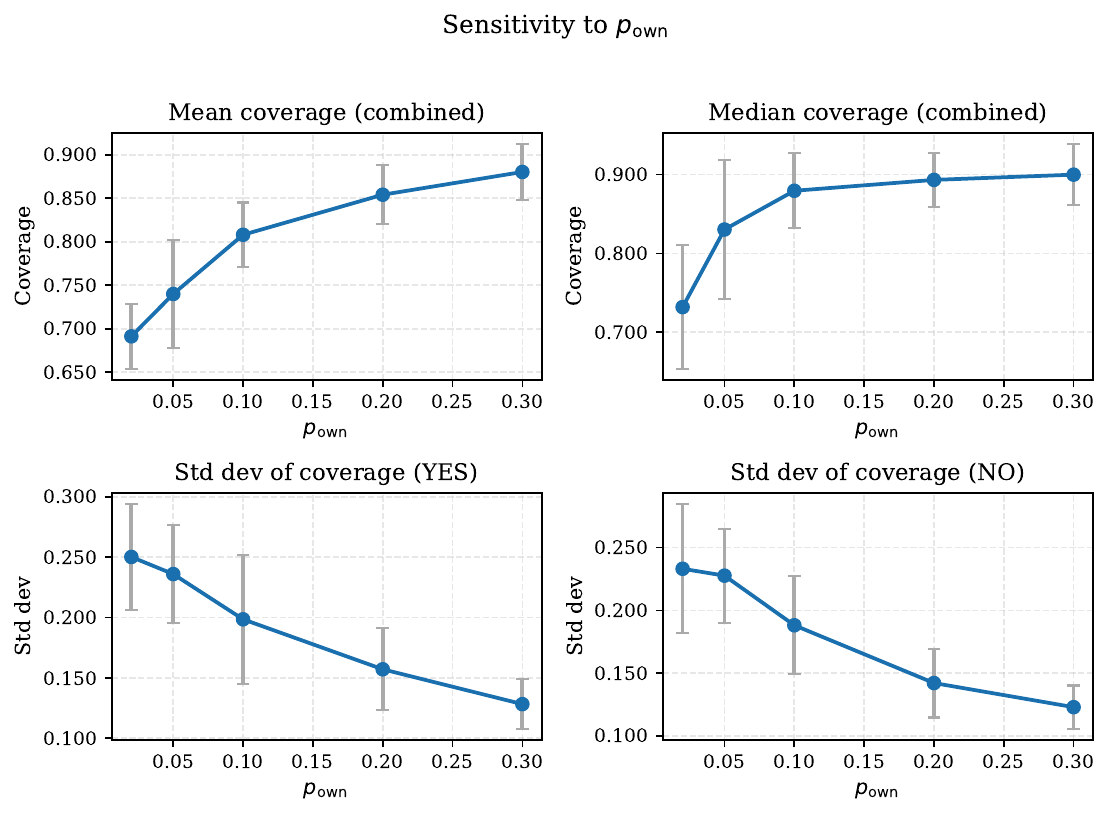}
  \caption{
  (BRA topic.) Mean combined recommendation coverage
    of the endorsement-ranked top-$K$ list as a function of
    creativity rate $p_{\mathrm{own}}$, averaged over 10 random seeds
    ($N=1{,}000$, $K=20$, $p_{\mathrm{links}}=0.60$). Error bars
    indicate $\pm 1$ standard deviation across seeds. Coverage increases
    monotonically with $p_{\mathrm{own}}$ and the across-seed variance
    decreases, reflecting a larger and more stable corpus. The two lower
    panels report the within-run coverage spread on each side.}
  \label{fig:pown_coverage}
\end{figure}

Figure~\ref{fig:pown_coverage} shows mean combined coverage as a function
of $p_{\mathrm{own}}$, averaged across 10 seeds. The relationship is
strongly monotone: 
coverage rises from $0.693 \pm 0.037$ at
$p_{\mathrm{own}} = 0.02$ to $0.881 \pm 0.032$ at
$p_{\mathrm{own}} = 0.30$, a $19$ percentage-point improvement; UBI traces the
same curve ($0.687 \to 0.877$). Creativity rate, recommendation budget $K$,
and population size $N$ are the three substantial coverage levers (see below).

The mechanism is straightforward: higher $p_{\mathrm{own}}$ generates
more shareholder-authored justifications, diversifying the corpus. At
$p_{\mathrm{own}} = 0.02$ only approximately $21$ own justifications
are contributed per run (10 YES, 11 NO), a sparse corpus from which
the endorsement-ranked top-$K$ list covers only a subset of the 30
tags per side; at $p_{\mathrm{own}} = 0.30$ approximately $300$ are
contributed and the corpus covers all tags many times over.
The
across-seed standard deviation decreases mildly with $p_{\mathrm{own}}$ ($0.037$
at $0.02$ to $0.032$ at $0.30$), the corpus reaches the full 30/30-tag vocabulary
once $p_{\mathrm{own}}\!\geq\!0.10$, and tag richness stays balanced between the
two sides, consistent with the symmetric uniform opinion distribution.

\paragraph{Early-Arrival Penalty.}
A critical concern is the \emph{early-arrival penalty}: shareholders who
vote before the corpus matures receive less comprehensive recommendations
than late-arriving shareholders. At $p_{\mathrm{own}} = 0.02$, the
typical corpus reaches full 30/30-tag coverage only after approximately
800 of the 1,000 shareholders have voted, meaning that the earliest 80\%
of voters are exposed to incomplete argument spaces. At
$p_{\mathrm{own}} = 0.30$, full coverage is typically achieved after
approximately 150 shareholders, benefiting the vast majority of the
electorate.

YES-side and NO-side mean coverage rise together and stay closely
matched across $p_{\mathrm{own}}$, confirming the symmetry of the two argument
spaces under the uniform opinion distribution.

\subsection{Effect of Recommendation Size ($K$)}
\label{sec:res_k}

\begin{figure}[t]
  \centering
  \includegraphics[width=\linewidth]{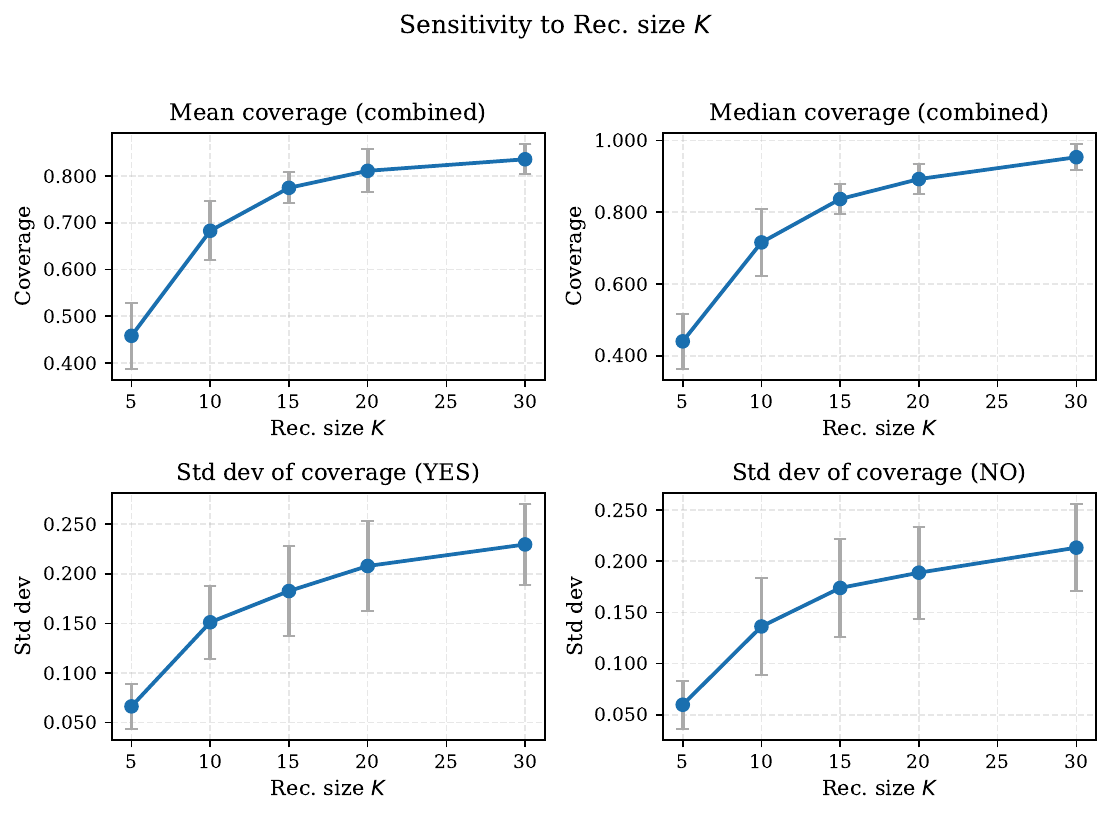}
  \caption{
  (BRA topic.) Mean combined recommendation coverage of
    the endorsement-ranked top-$K$ list as a function of
    recommendation size $K$, averaged over 10 random seeds ($N=1{,}000$,
    $p_{\mathrm{own}}=0.10$, $p_{\mathrm{links}}=0.60$). Coverage rises
    sharply from $K=5$ to $K=15$ and then climbs only marginally, so
    justifications beyond $K=15$ add little coverage in the baseline corpus.}
  \label{fig:recsize}
\end{figure}

Figure~\ref{fig:recsize} shows mean combined coverage as a function of
recommendation size $K \in \{5, 10, 15, 20, 30\}$. Coverage
rises steeply from $K = 5$ to $K = 15$ and then continues to climb with
sharply diminishing returns.

{Quantitatively (BRA; UBI within $\pm0.01$), mean combined coverage
runs $0.463\pm0.071$ at $K=5$, $0.684\pm0.063$ at $K=10$, $0.776\pm0.033$ at
$K=15$, $0.811\pm0.046$ at $K=20$, and $0.836\pm0.032$ at $K=30$.}

{The steep rise to $K=15$ arises because the baseline corpus,
with approximately $97$ total own justifications (49 YES, 48 NO), needs roughly
15 carefully ranked justifications to surface most of the 30 tags on each side.
Unlike the greedy oracle, the endorsement ranking is not tag-aware, so a few
additional slots beyond $K=15$ keep adding tags as lower-endorsement but
tag-novel justifications enter the top-$K$ -- hence the small but non-zero gains
from $K=20$ to $K=30$.}

This finding has a practical implication: for the baseline corpus size,
most of the attainable coverage is captured by the first $15$
justifications per side, and the marginal epistemic value of additional slots is
small. System designers should
calibrate $K$ to the expected corpus size at the typical voter's arrival
time, not to the final corpus size.

{The across-seed standard deviation of coverage stays in a narrow
band for $K \geq 15$ ($\sigma_c \approx 0.03$--$0.05$). }
{The unique-tag and relation counts are essentially unchanged across
$K$ values; these quantities are determined by $p_{\mathrm{own}}$ and
$p_{\mathrm{links}}$, not by the recommendation budget, as expected, the budget
affects only the quality of each voter's information exposure, not the rate of
corpus growth.}

\subsection{Effect of Link Probability ($p_{\mathrm{links}}$)}
\label{sec:res_links}

The link probability $p_{\mathrm{links}}$ controls the rate at which
shareholders submit attack and enhance relations when they hold a
justification. Unlike $p_{\mathrm{own}}$ and $K$, varying
$p_{\mathrm{links}}$ has negligible impact on recommendation coverage:
mean combined
coverage remains essentially constant at $\bar{c} \approx 0.81$ (range
$0.800$--$0.814$) across all
values $p_{\mathrm{links}} \in \{0.20, 0.40, 0.60, 0.80, 1.00\}$, even as the
total relation count grows roughly linearly (from $278$ edges at
$p_{\mathrm{links}}=0.20$ to $1{,}012$ at $1.00$) and tag richness stays at the
full 30/30 vocabulary.

{This result reveals a near-decoupling between argumentation-graph
density and coverage. Although the endorsement ranking of Eq.~\ref{eq:score}
does read the relation graph, in an \emph{honest} electorate the author-count
weights are small and additional enhance and attack edges accrue roughly
symmetrically across the pool, so raising $p_{\mathrm{links}}$ shifts the
relative ranking (and hence coverage) only marginally. The graph still
enriches the deliberative record and powers the transitive argumentation
coverage metric. This near-decoupling holds only while relations are
unconcentrated; Section~\ref{sec:res_attackers} shows that a coalition which
deliberately concentrates relations on a single hub can move coverage sharply.}

At $p_{\mathrm{links}} = 1.00$, every shareholder who holds a
justification submits both an attack and an enhance relation, yielding
approximately $1{,}010$ total edges, about $1.5\times$ the number
at the baseline $p_{\mathrm{links}} = 0.60$ ($689$ edges). This dense graph is informative
for the transitive argumentation coverage metric $\rho(j)$: with more
edges, the argumentation graph becomes more connected and individual
justifications can reach larger fractions of the electorate transitively.

\subsection{Effect of Population Size ($N$)}
\label{sec:res_n}

{Across five population sizes ($N = 200$, $500$, $1{,}000$,
$2{,}000$, $5{,}000$, log-scaled axis), mean combined coverage improves
substantially and monotonically: $0.693\pm0.051$, $0.745\pm0.055$,
$0.803\pm0.037$, $0.850\pm0.028$, and $0.898\pm0.028$ respectively.}

The improvement with $N$ arises because a larger population, at fixed
$p_{\mathrm{own}} = 0.10$, generates a larger own-justification corpus
(approximately $95$ justifications at $N = 1{,}000$ vs.\ approximately
$460$ at $N = 5{,}000$) and reaches full tag coverage earlier in the
sequence: at $N = 200$ the corpus may not mature before all shareholders
have voted, whereas at $N = 5{,}000$ it matures after approximately 400
shareholders, leaving the remaining 4,600 with near-comprehensive lists.
The across-seed variance also decreases with $N$ ($\sigma_c \approx
0.05$ at $N=200$ to $0.03$ at $N=5{,}000$), and both sides improve uniformly with
the coverage gap between them staying small throughout, reflecting the symmetric
uniform opinion distribution.

\subsection{Summary of Parameter Effects}

Table~\ref{tab:allresults} consolidates the mean combined coverage for
all parameter settings studied. The dominant factors are $p_{\mathrm{own}}$
and $N$ (which jointly determine corpus richness) and $K$ (which
determines how much of the corpus each voter can see). The link
probability $p_{\mathrm{links}}$ is orthogonal to coverage.

\begin{table}[t]
\centering
\caption{{Mean combined coverage $\bar{c}$ (and standard deviation
  $\sigma$ across 10 seeds) for all parameter settings, for both topics under the
  weighted relation mode. Baseline
  values are in bold. For $p_{\mathrm{own}}$ experiments,
  $p_{\mathrm{existing}} = 0.50$ and $p_{\mathrm{none}}$ is adjusted;
  all other parameters held at baseline.}}
\label{tab:allresults}
\setlength{\tabcolsep}{5pt}
\begin{tabular}{ll rr rr}
\toprule
& & \multicolumn{2}{c}{\textbf{BRA}} & \multicolumn{2}{c}{\textbf{UBI}} \\
\cmidrule(lr){3-4}\cmidrule(lr){5-6}
\textbf{Parameter} & \textbf{Value} & $\bar{c}$ & $\sigma$ & $\bar{c}$ & $\sigma$ \\
\midrule
\multirow{5}{*}{$p_{\mathrm{own}}$}
  & 0.02              & 0.693 & 0.037 & 0.687 & 0.069 \\
  & 0.05              & 0.741 & 0.062 & 0.743 & 0.060 \\
  & \textbf{0.10}     & \textbf{0.808} & \textbf{0.037} & \textbf{0.790} & \textbf{0.031} \\
  & 0.20              & 0.854 & 0.034 & 0.845 & 0.044 \\
  & 0.30              & 0.881 & 0.032 & 0.877 & 0.033 \\
\midrule
\multirow{5}{*}{$K$}
  & 5                 & 0.463 & 0.071 & 0.473 & 0.123 \\
  & 10                & 0.684 & 0.063 & 0.664 & 0.056 \\
  & 15                & 0.776 & 0.033 & 0.772 & 0.033 \\
  & \textbf{20}       & \textbf{0.811} & \textbf{0.046} & \textbf{0.806} & \textbf{0.029} \\
  & 30                & 0.836 & 0.032 & 0.833 & 0.020 \\
\midrule
\multirow{5}{*}{$p_{\mathrm{links}}$}
  & 0.20              & 0.814 & 0.031 & 0.798 & 0.035 \\
  & 0.40              & 0.808 & 0.040 & 0.799 & 0.033 \\
  & \textbf{0.60}     & \textbf{0.800} & \textbf{0.052} & \textbf{0.806} & \textbf{0.030} \\
  & 0.80              & 0.803 & 0.049 & 0.800 & 0.027 \\
  & 1.00              & 0.809 & 0.038 & 0.804 & 0.037 \\
\midrule
\multirow{5}{*}{$N$}
  & 200               & 0.693 & 0.051 & 0.682 & 0.071 \\
  & 500               & 0.745 & 0.055 & 0.733 & 0.048 \\
  & \textbf{1{,}000}  & \textbf{0.803} & \textbf{0.037} & \textbf{0.796} & \textbf{0.028} \\
  & 2{,}000           & 0.850 & 0.028 & 0.858 & 0.036 \\
  & 5{,}000           & 0.898 & 0.028 & 0.902 & 0.027 \\
\bottomrule
\end{tabular}
\end{table}

\subsection{{Adversarial Robustness}}
\label{sec:res_attackers}

{We now turn from honest electorates to the adversarial model of
Section~\ref{sec:setup_attackers}, in which a coalition of $0$--$25\%$ of
shareholders manipulates the relation graph that feeds the endorsement score
(Eq.~\ref{eq:score}). Because every identity is authenticated, the coalition
cannot inflate vote counts or forge endorsements; it can only choose where to
point its enhance and attack relations. The two strategies have opposite
consequences for coverage.}

\paragraph{{Hub-riding leaves coverage intact.}}
{When attackers merely launder the hub's endorsements into their
own legitimately-tagged justifications, coverage does not move: across all
attacker counts and both topics, every change from baseline is statistically
non-significant ($n=200$ seeds, all $p>0.05$), and the uniform-versus-weighted
gap stays below $0.002$. Coverage is a \emph{set-level} property -- promoting one
genuinely-tagged justification merely substitutes one covering item for another,
so the union of tags is preserved. Hub-riding distorts \emph{rankings} (its
proximate goal) without degrading \emph{coverage}.}

\begin{figure}[t]
  \centering
  \includegraphics[width=\linewidth]{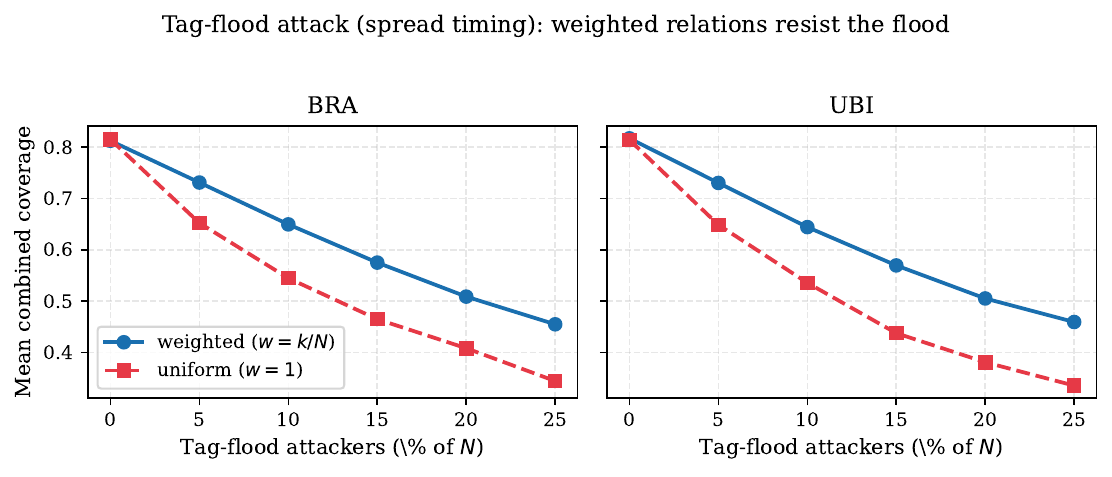}
  \caption{{Tag-flood attack under the default spread timing:
    mean combined coverage (200 seeds) vs.\ the fraction of tag-flood attackers,
    for BRA (left) and UBI (right). In both topics weighted relations
    ($w=k/N$, solid) preserve substantially more coverage than uniform weights
    ($w=1$, dashed); the gap widens to $\approx 0.11$--$0.12$ at $25\%$
    attackers.}}
  \label{fig:tagflood}
\end{figure}

\paragraph{{Tag-flood collapses coverage, and weighting is the
defence.}}
{The picture inverts when the coalition floods the top-$K$ with
tag-homogeneous clones that all ride the same hubs (Figure~\ref{fig:tagflood}).
Now each promoted clone carries the \emph{same} two poison tags, so filling
recommendation slots with clones evicts tag-diverse items and coverage falls
steeply -- from $\approx 0.81$ to $0.46$ (weighted) or $0.34$ (uniform) at $25\%$
attackers, every cell highly significant ($p<.001$). Crucially, the
author-count weighting throttles each solo clone$\to$hub edge to
$w=k/N$, starving clones of inherited endorsement mass, whereas uniform weights
hand every clone the hub's full score. Weighted relations therefore resist the
flood markedly better at every attacker count (Table~\ref{tab:attackers}; e.g.\
$0.455$ vs.\ $0.344$ for BRA and $0.459$ vs.\ $0.335$ for UBI at $25\%$). This is
the central practical finding of our adversarial study: non-uniform,
author-normalised relation weights are not a stylistic choice but a manipulation
defence.}

\begin{table}[t]
\centering
\caption{{Tag-flood attack (spread timing): mean combined coverage
  vs.\ attacker ratio, weighted vs.\ uniform relation modes, 200 seeds per cell.
  Hub-riding (not shown) produces no significant change at any ratio.}}
\label{tab:attackers}
\setlength{\tabcolsep}{6pt}
\begin{tabular}{r rr rr}
\toprule
& \multicolumn{2}{c}{\textbf{BRA}} & \multicolumn{2}{c}{\textbf{UBI}} \\
\cmidrule(lr){2-3}\cmidrule(lr){4-5}
\textbf{Attackers} & weighted & uniform & weighted & uniform \\
\midrule
{0\%}  & 0.812 & 0.815 & 0.817 & 0.814 \\
5\%  & 0.731 & 0.652 & 0.730 & 0.649 \\
10\% & 0.650 & 0.545 & 0.644 & 0.536 \\
15\% & 0.575 & 0.465 & 0.570 & 0.437 \\
20\% & 0.509 & 0.408 & 0.505 & 0.381 \\
25\% & 0.455 & 0.344 & 0.459 & 0.335 \\
\bottomrule
\end{tabular}
\end{table}

\section{Discussion}
\label{sec:discussion}

{The sharply diminishing returns of coverage beyond $K = 15$ under
baseline conditions reveal that the recommendation budget has a soft ceiling
determined by the effective vocabulary of the corpus at the time of voting. Most
attainable coverage is captured by the first $15$ justifications; because the
live recommender ranks by observable endorsements rather than tag content,
however, a few further slots keep admitting tag-novel items, so the ceiling is
approached gradually rather than hit exactly.} This has direct practical implications
for user interface design: presenting more than 15 justifications per
side yields only marginal epistemic benefit while adding
cognitive load. A dynamic $K$ that tracks the corpus-cover number would be more
efficient, but its implementation requires knowledge of the minimum set
cover, itself NP-hard.
The greedy oracle's
approximation guarantee provides a practical upper bound on the
achievable coverage for any fixed $K$.

\subsection{Diversity-Coverage Trade-off}

The strong positive effect of $p_{\mathrm{own}}$ on coverage identifies
shareholder creativity as the primary driver of epistemic quality, but it
carries a trade-off: higher $p_{\mathrm{own}}$ also grows the corpus with
increasingly tag-redundant justifications, so the marginal coverage gain of each
new entry decreases as the corpus matures---a concavity consistent with the
submodular structure of the coverage function. This motivates a \emph{novelty
filter} that declines to recommend (or merges) a new justification whose tag set
is already fully represented, reducing corpus bloat while preserving coverage.
Such a filter is also the natural defence against the tag-flood attack of
Section~\ref{sec:res_attackers}, which exploits precisely the absence of a
tag-novelty check in the endorsement ranking.

\subsection{Scalability: Coverage Improves with Population}

The positive dependence of coverage on $N$ is reassuring: larger
deliberative polls produce better-informed electorates, provided
$p_{\mathrm{own}}$ is held constant. However, the improvement is
sublinear—{increasing $N$ fivefold from 1,000 to 5,000 improves mean
BRA combined coverage by roughly 10 percentage points (from 0.803 to 0.898),
while a fivefold increase in $p_{\mathrm{own}}$ from 0.02 to 0.10
yields a comparable 11-point improvement (0.693 to 0.808) at a far smaller
participation cost}. Increasing $p_{\mathrm{own}}$ is therefore
a more efficient lever for improving coverage than simply recruiting
more participants.

The cross-seed variance analysis also reveals that smaller polls are
more susceptible to the vagaries of random shareholder orderings. This
suggests that deliberative platforms serving small communities ($N < 500$)
should seed the justification pool more aggressively with curated
system-generated content to compensate for the limited organic corpus.

\subsection{Limitations and Threats to Validity}

\paragraph{Synthetic text and tags.} The reason tags in \ABAS{} are
hand-crafted and assigned to justifications by a rule-based text
generator. Real deliberative corpora exhibit messier argument structures,
overlapping semantics, and idiosyncratic phrasing that may affect
coverage metrics.

\paragraph{No real user preferences.} The TF-IDF similarity mechanism
is a proxy for user affinity. Real voters may prefer short, emotionally
resonant arguments over topically diverse but dry enumeration. Survey
data or A/B testing on user engagement would be needed to calibrate the
recommendation mechanism to real-world behaviour.

\paragraph{No LLM integration.} The text generator produces deterministic
templates rather than LLM-generated prose. Integration of a language
model for justification generation would improve ecological validity but
would introduce stochasticity, latency, and cost.

\paragraph{{Bounded adversarial repertoire.}} {Because the electorate is
authenticated, Sybil attacks are impossible and every agent votes sincerely
from its latent opinion; the only gameable surface is the relation graph.
We model two strategies on that surface (hub-riding and tag-flood) under three
timing regimes. These do not exhaust the space of relation-level manipulations:
collusive multi-hub coordination, adaptive attackers that observe the live
recommendation list, and mixed honest/adversarial populations remain open.
Our results also show that the first-mover cascade (early timing) is a threat
that authentication and relation weighting alone do not neutralise, motivating
recommendation-level defences such as diversity-aware or novelty-penalised ranking.}

\paragraph{One-factor-at-a-time design.} Our sensitivity analysis does
not capture interactions between parameters. For example, the optimal
$K$ may depend on $p_{\mathrm{own}}$: if $p_{\mathrm{own}}$ is high
enough that the corpus matures very quickly, a smaller $K$ may suffice.
A full factorial design with 10 seeds would require $5^4 \times 10 =
6{,}250$ runs, which is computationally feasible but was outside the
scope of the present study.

\section{Conclusions and Future Work}
\label{sec:conclusion}

This paper has introduced \ABAS{},
a simulator for
deliberative polling grounded in the \BAPDF{} bipolar argumentation
framework. 
\ABAS{} models $N$ autonomous shareholder agents who sequentially vote,
author or reuse justifications, and submit attack and enhance relations,
with all state persisted in a
database. The simulator
serves each voter a recommendation list ranked by observable endorsement
score (a reversed-PageRank over weighted attack/enhance relations), and
evaluates the informational completeness of that list against a greedy
$K$-coverage oracle, which approximates the NP-hard Subsuming Justification
Problem via a formally defined coverage metric.

{We conducted 400 controlled experiments (5 parameter values $\times$ 4
parameters $\times$ 10 seeds, across the BRA and UBI topics) at $N = 1{,}000$
shareholders, plus dedicated adversarial sweeps, and characterised
the effects of four key parameters on mean combined coverage.}
Our principal findings are:
\begin{enumerate}[leftmargin=*,itemsep=2pt]
  \item \textbf{Creativity rate drives coverage.} $p_{\mathrm{own}}$ is
    the dominant determinant of recommendation coverage, with a roughly 19
    percentage-point improvement (0.69 to 0.88 for BRA) from the lowest to the highest value
    tested. Encouraging shareholders to author own justifications is the
    most effective intervention for improving epistemic quality.
  \item \textbf{{Coverage shows diminishing returns in $K$.}} {Marginal coverage gains
    from increasing the recommendation budget shrink steadily but, under the
    endorsement-ranked live recommender, coverage keeps rising past $K = 15$
    (0.78 at $K{=}15$ to 0.84 at $K{=}30$ for BRA) rather than saturating}. Recommendation interfaces should calibrate $K$ to the
    expected corpus maturity at the time of voting.
  \item \textbf{Graph density is orthogonal to coverage in the honest regime.} Argumentation
    link probability has negligible effect on recommendation coverage when agents behave honestly, revealing
    an empirical decoupling between the two objectives that nonetheless breaks
    when adversaries concentrate relation mass (see below).
  \item \textbf{Larger populations yield better coverage.} Coverage
    improves sublinearly with $N$, and run-to-run variance decreases,
    because larger populations generate richer corpora that mature faster
    relative to the electorate size.
  \item \textbf{{Tag homogeneity, not relation volume, is the attack vector.}} {With an
    authenticated electorate Sybil attacks are impossible, so adversaries can only
    game the relation graph. Hub-riding leaves coverage intact, but tag-flood clones
    collapse it (BRA 0.81 to 0.46 at 25\% attackers under spread timing), and
    author-count relation weighting consistently throttles the damage relative to
    uniform weighting (0.46 vs 0.34).}
  \item \textbf{Early-arrival penalty is structurally unavoidable} in
    sequential deliberation with a single round and is most severe when $p_{\mathrm{own}}$
    and $N$ are small. Seeding the pool with curated system justifications
    partially mitigates this penalty, but raises fairness and ethical concerns.
    Incentivising a second visit per shareholder
    might be more ethical but challenging.
\end{enumerate}
The present work opens several research directions:

\begin{enumerate}[leftmargin=*,itemsep=2pt]
  \item \textbf{LLM-generated justifications.} Replacing the rule-based
    text generator with a prompted large language model would improve
    ecological validity and allow testing on real argument corpora, the
    challenge being to keep a controlled, auditable reason-tag vocabulary
    under free-form generative text.
  \item \textbf{{Adaptive and collusive adversaries.}} {Modelling attackers
    that observe the live recommendation list and coordinate across multiple hubs
    would stress-test the endorsement-ranking recommender beyond the static,
    single-strategy regime and motivate diversity-aware ranking defences against
    the first-mover cascade.}
  \item \textbf{Weighted shareholders.} Implementing non-uniform $\VA$
    (e.g.\ differing domain expertise or stake) and studying its effect on
    coverage fairness across stakeholder groups.
  \item \textbf{Corpus validation and deployment.} Applying \ABAS{} with reason
    tags from an argument-mining pipeline to real deliberative corpora
    (parliamentary transcripts) and validating the coverage metric
    against expert assessments.
  \item \textbf{Multi-topic and multi-round dynamics.} Extending \BAPDF{} to
    co-occurring propositions with shared shareholders, and to longitudinal
    deliberation in which the corpus evolves and voter opinions update across
    rounds, enabling analysis of cross-topic transfer and opinion dynamics.
  \item \textbf{Bias, fairness, and coalition structure.} Analysing whether
    the endorsement-ranking recommender disadvantages minority
    viewpoints (whose tags appear in fewer justifications), developing
    fairness-aware variants that guarantee minimum coverage, and using the
    argumentation graph to detect emerging coalitions and factions.
\end{enumerate}

\ABAS{} provides a formal, reproducible, and extensible foundation for
the computational study of deliberative democracy at scale. By
operationalising the \BAPDF{} framework and measuring coverage in a
principled simulation environment, it bridges the theoretical richness
of formal argumentation with the practical demands of AI-assisted civic
deliberation. We release the simulator, experimental scripts, and all
results as open-source software to support reproducibility and to invite
community extensions.

\paragraph{Acknowledgements.}
AI help was used for coding and for drawing images. It was also asked to propose versions of refined expressions. The final text represents the expression choice, work, and ideas of the authors. We use Claude, Gemini, OpenAI, and DeepSeek.

\let\oldthebibliography\thebibliography
\renewcommand{\thebibliography}[1]{%
  \oldthebibliography{#1}%
  \setlength{\itemsep}{0pt}%
  \setlength{\parskip}{0pt}}
\bibliographystyle{splncs04}
\bibliography{refs}

\end{document}